\documentclass[a4paper,11pt]{article}
\usepackage{jheppub} 

\usepackage[english]{babel}
\usepackage{amsmath,amssymb,amsfonts, bm,bbm,slashed, subdepth}
\usepackage{graphicx}
\usepackage[sort&compress]{natbib}
\usepackage{xcolor}
\usepackage[normalem]{ulem}
\usepackage{hyperref}
\usepackage{cleveref}
\definecolor{red}{rgb}{1.0, 0, 0}
\usepackage{enumerate}
\usepackage{epsfig, subfigure}
\usepackage{setspace}
\usepackage{booktabs, tabularx}
\usepackage{units}
\usepackage{placeins}
\usepackage{multirow}
\usepackage{mathtools}

\usepackage{graphicx}
\usepackage{dcolumn}
\usepackage{bm}

\newcommand{\bmat}{\left(\begin{array}}
	\newcommand{\emat}{\end{array}\right)}
\newcommand{\beq}{\begin{equation}}
	\newcommand{\eeq}{\end{equation}}
\def\beq{\begin{equation}}
	\def\eeq{\end{equation}}

\def\beqx{\begin{displaymath}}
	\def\eeqx{\end{displaymath}}

\def\beqa{\begin{eqnarray}}
	\def\eeqa{\end{eqnarray}}

\newcommand{\drawsquare}[2]{\hbox{%
\rule{#2pt}{#1pt}\hskip-#2pt
\rule{#1pt}{#2pt}\hskip-#1pt
\rule[#1pt]{#1pt}{#2pt}}\rule[#1pt]{#2pt}{#2pt}\hskip-#2pt
\rule{#2pt}{#1pt}}

\def\bo{{\raise-.3ex\hbox{\large$\Box$}}}               
\def\face{{\raise.2ex\hbox{$\displaystyle \bigodot$}\mskip-2.2mu \llap {$\ddot
			\smile$}}}                                      

\def\yzero{\smash{\hbox{$y\kern-4pt\raise1pt\hbox{${}^\circ$}$}}}

\def\-{\hphantom{-}}
\def\ov{\overline}
\def\s2{\frac{1}{\sqrt2}}

\def\beq{\begin{equation}}
	\def\eeq{\end{equation}}
\def\beqa{\begin{eqnarray}}
	\def\eeqa{\end{eqnarray}}

\def\IF{\relax{\rm I\kern-.18em F}}
\def\II{\relax{\rm I\kern-.18em I}}
\def\IP{\relax{\rm I\kern-.18em P}}

\def\Dsl{\,\raise.15ex\hbox{/}\mkern-13.5mu D} 

\def\IC{\bf C}
\def\IZ{\bf Z}

\def\z2z2{$\IC^3/(\IZ_2\times\IZ_2)$}

\newcommand{\fund}{\raisebox{-.5pt}{\drawsquare{6.5}{0.4}}}
\newcommand{\Ysymm}{\raisebox{-.5pt}{\drawsquare{6.5}{0.4}}\hskip-0.4pt%
        \raisebox{-.5pt}{\drawsquare{6.5}{0.4}}}
\newcommand{\Yasymm}{\raisebox{-3.5pt}{\drawsquare{6.5}{0.4}}\hskip-6.9pt%
        \raisebox{3pt}{\drawsquare{6.5}{0.4}}}
\newcommand{\antifund}{\overline{\fund}}

 \renewcommand{\thetable}{\arabic{table}}

\begin{document}

\title{The Final Model Building for the Supersymmetric Pati-Salam Models from Intersecting D6-Branes}

\author[a,b,1]{Weikun He}
\author[c,d,2]{Tianjun Li}
\author[a,3]{Rui Sun\footnote{Corresponding author.}}
\author[e,4]{Lina Wu}

\affiliation[a]{Korea Institute for Advanced Study,\\
	85 Hoegiro, Dongdaemun-Gu, Seoul 02455, Korea}
\affiliation[b]{Academy of Mathematics and Systems Science,\\
	Beijing 100190, P. R. China}
\affiliation[c]{CAS Key Laboratory of Theoretical Physics, Institute of Theoretical Physics,\\
	Chinese Academy of Sciences, Beijing 100190, P. R. China}
\affiliation[d]{School of Physical Sciences, University of Chinese Academy of Sciences,\\
	No.19A Yuquan Road, Beijing 100049, P. R. China}
\affiliation[e]{School of Sciences, Xi'an Technological University, 
	Xi'an 710021,  P. R. China}

\emailAdd{heweikun@amss.ac.cn}
\emailAdd{tli@itp.ac.cn}
\emailAdd{sunrui@kias.re.kr}
\emailAdd{wulina@xatu.edu.cn}

\abstract{
All the possible three-family ${\cal N}=1$ supersymmetric Pati-Salam models constructed with intersecting D6-branes from Type IIA orientifolds on  $T^6/(\mathbb{Z}_2\times \mathbb{Z}_2)$  are  recently presented in arXiv: 2112.09632. Taking  models with largest wrapping number $5$ and approximate gauge coupling unification at GUT scale as examples, we show string scale gauge coupling unification can be realized through two-loop renormalization group equation running by introducing seven pairs of vector-like particles from ${\cal N}=2$ sector. The number of these introduced vector-like particles are fully determined by the brane intersection numbers while there are two D6-brane parallel to each other along one two-torus. We expect this will solve the gauge coupling unification problem in the generic intersecting brane worlds by introducing vector-like particles that naturally included in the ${\cal N}=2$ sector. 
}

\maketitle

\section{Introduction} 
 How to construct the realistic string models which are consistent with the low energy phenomenology is a great
challenge question. Among various string models, the Pati-Salam model building in Type IIA theory on 
$T^6/(\mathbb{Z}_2\times \mathbb{Z}_2)$  orientifold 
with intersecting D6-branes~\cite{lpt,magnetised,MCJW, Ibanez,MKRR,urangac,afiru,Uranga,imr,bgkl,bkl,CSU2, Blumenhagen:2003jy,CPS,Cvetic:2004ui,Blumenhagen:2005mu,Blumenhagen:2006ci,CLW,blumrecent,Honecker,Cvetic:2004nk,Chen:2005mj,Chen:2007px,Chen:2007zu}  
is very interesting, although the absence of gauge coupling unification at the string scale might be a generic problem.
Recently,  some concrete supersymmetric Pati-Salam models have been constructed~\cite{Li:2019nvi, Li:2021}, 
and such D-brane models have been investigated with more powerful reinforcement machine learning methods and genetic algorithms
 as shown in Refs.~\cite{Jim:brane, Loges:2021hvn}.

In~\cite{Li:2021-2}, we propose a systematic method for such kind of model building
and for the first time construct all the three-family ${\cal N}=1$ supersymmetric Pati-Salam models 
from intersecting D6-branes where the gauge symmetry can be broken down to the Standard Model (SM) 
gauge symmetry via the D-brane splitting and supersymmetry (SUSY) preserving Higgs mechanism.
We find that there are $202752$ models in total, and show that there are only 33 independent models 
with different gauge coupling relations at string scale after moduloing the equivalent relations.
Thus, for the string landscape study, we would like to emphasize that 
it is very important to consider the equivalent relations.
Howevere,  one and only one model has traditional gauge coupling unification at string scale. 
We also find that the largest absolute value of wrapping number is $5$.

In~\cite{Blumenhagen:2003jy},  additional 
vector-like matter was introduced to push GUT scale unification up to Planck scale for $SU(5)$ and $SO(10)$ GUT theories. 
Instead of studying pushing up the unification scale,  we will study the methods to achieve string scale gauge unification from ${\cal N}=1$ supersymmetric Pati-Salam models without GUT scale unification.
Taking  models with largest wrapping number  $5$ and approximate gauge coupling unification as an examples, 
we show that the additional exotic particles can be decoupled and its gauge coupling relation 
can be realized at string scale via two-loop renormalization group equation running by introducing 
seven pairs of vector-like particles from ${\cal N}=2$ sector.
Therefore, 
although most of these Pati-Salam models do not directly have gauge coupling unification at string scale, 
their gauge coupling unification may indeed be achieved at string scale by introducing the vector-like particles 
from ${\cal N}=2$ sector. And then we solve the gauge coupling unification problem in the generic D-brane models.
Moreover, one can explain the SM fermion masses and mixings via the three-point and four point functions~\cite{LMSW}.
Furthermore, it seems to us that this brand new systematic method can be applied to the other intersecting D-brane model building as well.


\section{Review of Model Building}
First, we briefly review the basics of constructing supersymmetric Pati-Salam models on Type IIA $T^6/(\mathbb{Z}_2\times \mathbb{Z}_2)$  orientifolds with D6-branes intersecting at generic angles
as discussed in~\cite{CSU2, CPS}.
Consider the six-torus $T^{6}$ factorized as three two-tori  $T^{6} = T^{2} \times T^{2} \times T^{2}$. 
Let $z_i$, $i=1,\; 2,\; 3$ be respectively the complex coordinates for the $i$-th two-torus.
For the orbifold group $\mathbb{Z}_{2} \times \mathbb{Z}_{2}$, the generators $\theta$ and $\omega$ are associated with the twist vectors $(1/2,-1/2,0)$ and $(0,1/2,-1/2)$ respectively. They act on the complex coordinates $z_i$ as
\begin{equation}\label{orbifold}
    \begin{split}
    \theta: & (z_1,z_2,z_3) \mapsto (-z_1,-z_2,z_3)~,~  \\
    \omega: & (z_1,z_2,z_3) \mapsto (z_1,-z_2,-z_3)~.~  
    \end{split}
\end{equation}
The orientifold projection is implemented by gauging the $\Omega R$ symmetry, where $\Omega$ is world-sheet parity and $R$ acts on the complex coordinates as
\begin{equation}
    R: (z_1,z_2,z_3) \mapsto ({\ov z}_1,{\ov z}_2,{\ov
	z}_3)~.~\, 
\end{equation}
In total, we have four kinds of orientifold 6-planes (O6-planes) for the actions of $\Omega R$, $\Omega R\theta$, $\Omega R\omega$, and $\Omega R\theta\omega$ respectively~\cite{bkl, Chen:2007zu, CSU2,CPS}. 
Moreover, three stacks of $N_a$ D6-branes wrapping on the factorized three-cycles are introduced to cancel the RR charges of these O6-planes. 

There are two kinds of complex structures: rectangular and tilted, consistent with the orientifold projection. 
The homology classes of these three cycles wrapped by the D6-brane stack are of the form $n_a^i[a_i]+m_a^i[b_i]$ for a rectangular torus and $n_a^i[a'_i]+m_a^i[b_i]$ for a tilted torus, where $[a_i']=[a_i]+\frac{1}{2}[b_i]$.
A generic one cycle can be labeled by $(n_a^i,l_a^i)$ in terms of these wrapping numbers, with $l_{a}^{i}\equiv m_{a}^{i}$ and $l_{a}^{i}\equiv 2\tilde{m}_{a}^{i}=2m_{a}^{i}+n_{a}^{i}$ for a rectangular and tilted two-torus respectively. Note that $l_a^i-n_a^i$ is an even number for tilted two-tori.
We denote the wrapping number for stack $a$ of D6-branes along the cycle to be $(n_a^i,l_a^i)$, and their $\Omega R$ images ${a'}$ stack of $N_a$ D6-branes are with wrapping numbers $(n_a^i,-l_a^i)$. 
The homology three-cycles 
for stack $a$ of D6-branes and  its orientifold image  $a'$  can be expressed as
\begin{eqnarray} \label{homo}
&[\Pi_a]=\prod_{i=1}^{3}\left(n_{a}^{i}[a_i]+2^{-\beta_i}l_{a}^{i}[b_i]\right),\\ 
&[\Pi_{a'}]=\prod_{i=1}^{3} \left(n_{a}^{i}[a_i]-2^{-\beta_i}l_{a}^{i}[b_i]\right),
\end{eqnarray}
where $\beta_i=0$ if the $i$-th two-torus is rectangular and $\beta_i=1$ otherwise.
The homology three-cycles wrapped by the four O6-planes take the forms of 
\begin{equation}\label{orienticycles}
    \begin{split}
        &\Omega R: [\Pi_{\Omega R}]= 2^3 [a_1]\times[a_2]\times[a_3]~,~\,\\
		& \Omega R\omega: [\Pi_{\Omega	R\omega}]=-2^{3-\beta_2-\beta_3}[a_1]\times[b_2]\times[b_3]~,~\\
		&\Omega R\theta\omega: [\Pi_{\Omega	R\theta\omega}]=-2^{3-\beta_1-\beta_3}[b_1]\times[a_2]\times[b_3]~,~\\
		&\Omega R\theta:  [\Pi_{\Omega	R}]=-2^{3-\beta_1-\beta_2}[b_1]\times[b_2]\times[a_3]~.~  
    \end{split}
\end{equation}
The intersection numbers can be expressed in terms of the wrapping numbers as
\begin{eqnarray} \label{intersection}
	&I_{ab}=[\Pi_a][\Pi_b]=2^{-k}\prod_{i=1}^3(n_a^il_b^i-n_b^il_a^i)~,~\nonumber\\
	&I_{ab'}=[\Pi_a][\Pi_{b'}]=-2^{-k}\prod_{i=1}^3(n_{a}^il_b^i+n_b^il_a^i)~,~\nonumber\\
	&I_{aa'}=[\Pi_a][\Pi_{a'}]=-2^{3-k}\prod_{i=1}^3(n_a^il_a^i)~,~\\
	&I_{aO6}=[\Pi_a][\Pi_{O6}]=	\nonumber\\
	&2^{3-k}(-l_a^1l_a^2l_a^3	+l_a^1n_a^2n_a^3+n_a^1l_a^2n_a^3+n_a^1n_a^2l_a^3)\nonumber
\end{eqnarray}
where $k=\beta_1+\beta_2+\beta_3$ is the total number of tilted two-tori, while $[\Pi_{O6}]=[\Pi_{\Omega R}]+[\Pi_{\Omega R\omega}]+[\Pi_{\Omega R\theta\omega}]+[\Pi_{\Omega R\theta}]$ is the sum of the four O6-plane homology three-cycles.

The massless particle spectrum obtained from intersecting D6-branes at general angles can be expressed in terms of these intersection numbers as in  Table \ref{spectrum}.
Here, the representations refer to $U(N_a/2)$, e.g. when the intersecting D6-branes are of number $N_a=8, N_b=4, N_c=4$, this gives $U(N_x/2)$ gauge groups for $x=a, b, c$.
The gauge symmetry results from $\mathbb Z_2\times \mathbb Z_2$ orbifold projection, as discussed in~\cite{CSU2}.
The chiral supermultiplets contain both scalars and fermions in this supersymmetric constructions, while the positive intersection numbers refer to the left-handed chiral supermultiplets. 
To have three families of the SM fermions with quantum numbers are under $SU(4)_C\times SU(2)_L\times SU(2)_R$ gauge symmetries, we require the intersection numbers to satisfy
\begin{equation}
    \begin{split}\label{3gen}
      	I_{ab} + I_{ab'}~=~3~,~\\ 
		I_{ac} ~=~-3~,~ I_{ac'} ~=~0~,~\,  
    \end{split}
\end{equation}
or with $c$ and $c'$ exchange.
And then we can break the Pati-Salam gauge symmetry
 down to the SM gauge symmetry via the D-brane splitting 
and supersymmetry preserving Higgs mechanism~\cite{Cvetic:2004ui}.
\begin{table}[h]
	\caption{General massless particle spectrum for intersecting D6-branes at generic angles.}
	\label{spectrum}
	\renewcommand{\arraystretch}{1}
	\begin{center}
		\begin{tabular}{c|c}
			\hline {\bf Sector} & 
			{\bf Representation}
			\\
			\hline
			\hline
			$aa$   & $U(N_a/2)$ vector multiplet  \\
			& 3 adjoint chiral multiplets  \\
			\hline
			$ab+ba$   & $I_{ab}$ $(\fund_a,\antifund_b)$ fermions   \\
			\hline
			$ab'+b'a$ & $I_{ab'}$ $(\fund_a,\fund_b)$ fermions \\
			\hline $aa'+a'a$ &$\frac 12 (I_{aa'} - \frac 12 I_{a,O6})\;\;
			\Ysymm\;\;$ fermions \\
			& $\frac 12 (I_{aa'} + \frac 12 I_{a,O6}) \;\;
			\Yasymm\;\;$ fermions \\
			\hline
		\end{tabular}
	\end{center}
\end{table}

In addition to the three generation conditions Eq.\eqref{3gen}, there are two other main constraints on the four-dimensional $N=1$ supersymmetric model building, namely, the RR tadpole cancellation conditions and the $N=1$ supersymmetry preservation in four dimension~\cite{Uranga,imr,CSU2}.
In the brane constructions, D6-branes and the orientifold O6-planes are the sources of RR fields and restricted by the Gauss law in a compact space. Due to the conservations of the RR field flux lines, the RR charges from D6-branes  must cancel with it from the O6-planes, which results in the RR tadpole cancellations conditions
\begin{equation}  \label{tap1}
	\sum_a N_a [\Pi_a]+\sum_a N_a
	\left[\Pi_{a'}\right]-4[\Pi_{O6}]=0,
\end{equation} 
where the last term arises from the O6-planes are with $-4$ RR charges in D6-brane charge unit. 
To simplify the discussion on tadpole cancellation condition in the next section, we define
{\beq
\begin{array}{rrrr} 
	A_a \equiv -n_a^1n_a^2n_a^3,\,B_a \equiv n_a^1l_a^2l_a^3,\,
	C_a \equiv l_a^1n_a^2l_a^3,\, D_a \equiv l_a^1l_a^2n_a^3, \\
	\tilde{A}_a \equiv -l_a^1l_a^2l_a^3,\,  \tilde{B}_a \equiv
	l_a^1n_a^2n_a^3,  \,\tilde{C}_a \equiv n_a^1l_a^2n_a^3,\, 
	\tilde{D}_a \equiv n_a^1n_a^2l_a^3.\,
\end{array}
\label{variables}
\eeq }In order to cancel the RR tadpoles, we introduce D6-branes wrapping cycles along the orientifold planes as the so-called ``filler branes''.  They contributes to the RR tadpole cancellation conditions, and trivially satisfy the four-dimensional $N=1$ supersymmetry conditions. They are chosen  such that the tadpole conditions satisfy
\begin{equation}\label{tap2}
    \begin{split}
      &-2^k N^{(1)}+\sum_a N_a A_a=-2^k N^{(2)}+\sum_a N_a B_a= \\
	&-2^k N^{(3)}+\sum_a N_a C_a=-2^k N^{(4)}+\sum_a
	N_a D_a=-16,  
    \end{split}
\end{equation}
where $2 N^{(i)}$ is the number of filler branes wrapping along the $i$-th O6-plane. Moreover, these filler branes represent the $USp$ group and carry the wrapping numbers with one of the O6-planes as shown in Table~\ref{orientifold}.
\renewcommand{\arraystretch}{1}
\begin{table}[h]\small
	\caption{The wrapping numbers for four O6-planes.} 
	\begin{center}
		\begin{tabular}{c|c|c}
			\hline
			{\bf	Orientifold action} & 	{\bf O6-Plane }& $(n^1,l^1)\times (n^2,l^2)\times	(n^3,l^3)$\\
			\hline	\hline
			$\Omega R$& 1 & $(2^{\beta_1},0)\times (2^{\beta_2},0)\times
			(2^{\beta_3},0)$ \\
			\hline
			$\Omega R\omega$& 2& $(2^{\beta_1},0)\times (0,-2^{\beta_2})\times
			(0,2^{\beta_3})$ \\
			\hline
			$\Omega R\theta\omega$& 3 & $(0,-2^{\beta_1})\times
			(2^{\beta_2},0)\times
			(0,2^{\beta_3})$ \\
			\hline
			$\Omega R\theta$& 4 & $(0,-2^{\beta_1})\times (0,2^{\beta_2})\times
			(2^{\beta_3},0)$ \\
			\hline
		\end{tabular}
	\end{center}
	\label{orientifold}
\end{table}
Furthermore, the four-dimensional $N=1$ supersymmetry can be preserved with the orientation projection iff the rotation angle of any D6-brane with respect to the orientifold plane is an element of $SU(3)$, as shown in~\cite{bdl}. 
It follows that $\theta_1+\theta_2+\theta_3=0 $ mod $2\pi$, where $\theta_i$ is the angle between the $D6$-brane and the orientifold-plane in the $i$-th two-torus. 
The 4-dimensional N = 1 supersymmetry automatically survive the $\mathbb Z_2\times \mathbb Z_2$ orbifold projection~\cite{CPS}, and the SUSY conditions result in the equality and inequality conditions
\begin{equation}\label{eq:susy}
    \begin{split}
	x_A\tilde{A}_a+x_B\tilde{B}_a+x_C\tilde{C}_a+x_D\tilde{D}_a=0,\\
	A_a/x_A+B_a/x_B+C_a/x_C+D_a/x_D<0,
    \end{split}
\end{equation}
where $x_A=\lambda$,
$x_B=\lambda~2^{\beta_2+\beta3}/\chi_2\chi_3$,\; $x_C=\lambda
2^{\beta_1+\beta3}/\chi_1\chi_3,$\; $x_D=\lambda
2^{\beta_1+\beta2}/\chi_1\chi_2$, 
while $\chi_i=R^2_i/R^1_i$ represent the complex structure moduli for the $i$-th two-torus. The positive parameter $\lambda$  is introduced to put all the variables $A,\,B,\,C,\,D$ at equal footing.

\section{Search Method for Supersymmetric Pati-Salam Models}
From a mathematical point of view, the search for wrapping numbers satisfying all the conditions described in the last section is equivalent to solving a Diophantine equation.
In general, there does not exist any algorithm to solve such equation, as implied by the negative solution of Hilbert's tenth problem.
However, we devise an algorithm specially adapted to the form of our equations  for Supersymmetric Pati-Salam Models~\cite{Li:2021-2}.

Now we describe the main strategy of our algorithm.
First step: 
we enumerate all the possible combinations of the signs of the twelve wrapping number products $A_a,B_a,\dotsc,C_c,D_c$.
As observed previously in \cite{CPS}, because of the SUSY condition, $(A_a,B_a,C_a,D_a)$ contains three negative numbers and one positive number, or two negative numbers and two zeroes, or with  one negative and three zeros. The same applies to the stack $b$ and the stack $c$.

Second step: for each possibility listed in the first step, we append the twelve corresponding inequalities to the our system and try to solve the new system.
Inside the system, we look for the following  kinds of equations or subsystems
\begin{enumerate}
	\item Equation with only one non constant monome.
	
	\item A system of linear equations of full rank. 

	\item A system of linear inequalities which has finitely many integer solutions.

\end{enumerate}

When we find inside our system equations or subsystems of these kinds, we can solve them with respect to the variables appearing in them. After the resolution, the total number of unknown variables decreases. 
This might lead to branching into subcases because the solution might not be unique. But there are only {\bf  finitely} many subcases because we specifically looked for subsystems admitting finite integer solution sets.
Then, we repeat the procedure for each subcase.

Third step: we repeat the procedure described in the second step until there are no subsystem of the three kinds.
At this stage, each subcase either gives a solution or still has some unsolved varialbles (wrapping numbers). 
Now we make use of inequalities of degree larger than 1, such as those given by~Eq.\eqref{eq:susy}. 
By enumerating all the possible signs of the remaining variables, we can determine whether~Eq.\eqref{eq:susy} can be achieved.

Implementing this algorithm with the help of a computer program, we find that
 in total there are $202752$ models. Because each equivalent class has 6144 models,
we obtain that there are only 33 independent models 
with different gauge coupling relations at string scale after moduloing the equivalent relations.
Thus, for the string landscape study, it is very important to consider the equivalent relations.


\section{ Supersymmetric Pati-Salam Models }

As presented with details in \cite{Li:2021-2}, the full list of supersymmetric Pati-Salam models have $33$ different gauge coupling relations. 

First, there is one and only one inequivalent model with gauge coupling unification, {\it i.e.},
 the strong, weak and hypercharge gauge couplings $g^2_a,  g^2_b$ and $\frac{5}{3}g^2_Y$ satisfy $g^2_a=g^2_b=g^2_c=(\frac{5}{3}g^2_Y)=4\sqrt{\frac{2}{3}} \pi  e^{\phi ^4}$, where $\phi^4$ is the dilaton field.
Also, there are four confining $USp(2)$ gauge groups, and then we might
stabilize the moduli and break SUSY via gaugino condensation. 

Second, let us present three representative models for the following discussions of gauge coupling relations at string scale
in the next Section: Model 1 in  Table~\ref{tb:model1}  is a representative model with gauge coupling unification 
at  string scale; 
Model 2 in Table~\ref{tb:model2} and Model 3 in Table~\ref{tb:model3} are two kinds of 
models with the largest possible wrapping number 5, which are related to each other 
by exchanging $b-$ an $c-$stacks of D6-branes up to equivalent relations.
The gauge symmetry groups in Model 1 has four $USp(2)$ groups with negative $\beta$ functions,
so their moduli may be stabilized and  supersymmetry might be broken via gaugino condensations.
However, Models 2 and 3 contain two $USp(2)$ groups, and only one $USp(2)$ group has negative $\beta$ function. 

In Model 1, 
similar to~\cite{Chen:2007px}, the additional exotic particles can be decoupled.
In Models 2 and 3, similar to~\cite{Chen:2007px}, we can decouple  the additional exotic particles except 
the four chiral multiplets under $SU(4)_C$ anti-symmetric representation. And these four chiral multiplets can
be decoupled via instanton effects in principle~\cite{Blumenhagen:2006xt, Haack:2006cy, Florea:2006si}, which we will discuss in~\cite{Li:2021-3}. 

\begin{table}[h]\scriptsize
	\caption{D6-brane configurations and intersection numbers of Model 1, and its gauge coupling relation is $g^2_a=g^2_b=g^2_c=(\frac{5}{3}g^2_Y)=4 \sqrt{\frac{2}{3}} \pi  e^{\phi ^4}$.}
	\label{tb:model1}
	\begin{center}
		\begin{tabular}{|c|c|c|c|c|c|c|c|c|c|c|c|c|}
			\hline\rm{{Model 1}}  & \multicolumn{12}{c|}{$U(4)\times U(2)_L\times U(2)_R\times USp(2)^4 $}\\
			\hline \hline			\rm{stack} & $N$ & $(n^1,l^1)\times(n^2,l^2)\times(n^3,l^3)$ & $n_{S}$& $n_{A}$ & $b$ & $b'$ & $c$ & $c'$ & 1 & 2 & 3 & 4\\
			\hline
			$a$ & 8 & $(1,1)\times (1,0)\times (1,-1)$ & 0 & 0  & 3 & 0 & 0 & -3 & 0 & 1 & 0 & -1\\
			$b$ & 4 & $(-1,0)\times (-1,3)\times (1,1)$ & 2 & -2  & - & - & 0 & 0 & 0 & 0 & -3 & 1\\
			$c$ & 4 & $(0,1)\times (-1,3)\times (-1,1)$ & 2 & -2  & - & - & - & - & -3 & 1 & 0 & 0\\
			\hline
			1 & 2 & $(1, 0)\times (1, 0)\times (2, 0)$& \multicolumn{10}{c|}{$x_A = \frac{1}{3}x_B = x_C = \frac{1}{3}x_D$}\\
			2 & 2 & $(1, 0)\times (0, -1)\times (0, 2)$& \multicolumn{10}{c|}{$\beta^g_1=\beta^g_2=\beta^g_3=\beta^g_4=-3$}\\
			3 & 2 & $(0, -1)\times (1, 0)\times (0, 2)$& \multicolumn{10}{c|}{$\chi_1=1$, $\chi_2=\frac{1}{3}$, $\chi_3=2$}\\
			4 & 2 & $(0, -1)\times (0, 1)\times (2, 0)$& \multicolumn{10}{c|}{}\\
			\hline
		\end{tabular}
	\end{center}
\end{table}

\begin{table}[h]\scriptsize
	\caption{D6-brane configurations and intersection numbers of Model 2, and its gauge coupling relation is $g^2_a=\frac{7}{6}g^2_b=\frac{5}{6}g^2_c=\frac{25}{28}(\frac{5}{3}g^2_Y)=\frac{8}{27} 5^{3/4} \sqrt{7} \pi  e^{\phi ^4}$.}
	\label{tb:model2}
	\begin{center}
		\begin{tabular}{|c|c|c|c|c|c|c|c|c|c|c|}
			\hline\rm{Model 2}  & \multicolumn{10}{c|}{$U(4)\times U(2)_L\times U(2)_R\times USp(2)^2 $}\\
			\hline \hline			\rm{stack} & $N$ & $(n^1,l^1)\times(n^2,l^2)\times(n^3,l^3)$ & $n_{S}$& $n_{A}$ & $b$ & $b'$ & $c$ & $c'$ & 1 & 4\\
			\hline
			$a$ & 8 & $(1,-1)\times (-1,1)\times (1,-1)$ & 0 & 4  & 0 & 3 & 0 & -3 & -1 & 1\\
			$b$ & 4 & $(0,1)\times (-2,1)\times (-1,1)$ & -1 & 1  & - & - & 0 & -1 & -1 & 0\\
			$c$ & 4 & $(-1,0)\times (5,2)\times (-1,1)$ & 3 & -3  & - & - & - & - & 0 & -5\\
			\hline
			1 & 2 & $(1, 0)\times (1, 0)\times (2, 0)$& \multicolumn{8}{c|}{$x_A = 2x_B = \frac{14}{5}x_C = 7x_D$}\\
			4 & 2 & $(0, -1)\times (0, 1)\times (2, 0)$& \multicolumn{8}{c|}{$\beta^g_1=-3$, $\beta^g_4=1$}\\
			& & & \multicolumn{8}{c|}{$\chi_1=\frac{7}{\sqrt{5}}$, $\chi_2=\sqrt{5}$, $\chi_3=\frac{4}{\sqrt{5}}$}\\
			\hline
		\end{tabular}
	\end{center}
\end{table}

\begin{table}[!h]\scriptsize
	\caption{D6-brane configurations and intersection numbers of Model 3, and its gauge coupling relation is $g^2_a=\frac{5}{6}g^2_b=\frac{7}{6}g^2_c=\frac{35}{32}(\frac{5}{3}g^2_Y)=\frac{8}{27} 5^{3/4} \sqrt{7} \pi  e^{\phi ^4}$.}
	\label{tb:model3}
	\begin{center}
		\begin{tabular}{|c|c|c|c|c|c|c|c|c|c|c|}
			\hline\rm{Model 3} & \multicolumn{10}{c|}{$U(4)\times U(2)_L\times U(2)_R\times USp(2)^2 $}\\
			\hline \hline			\rm{stack} & $N$ & $(n^1,l^1)\times(n^2,l^2)\times(n^3,l^3)$ & $n_{S}$& $n_{A}$ & $b$ & $b'$ & $c$ & $c'$ & 1 & 4\\
			\hline
			$a$ & 8 & $(-1,-1)\times (1,1)\times (1,1)$ & 0 & -4  & 0 & 3 & 0 & -3 & 1 & -1\\
			$b$ & 4 & $(-5,2)\times (-1,0)\times (1,1)$ & -3 & 3  & - & - & 0 & 1 & 0 & 5\\
			$c$ & 4 & $(-2,-1)\times (0,1)\times (1,1)$ & 1 & -1  & - & - & - & - & 1 & 0\\
			\hline
			1 & 2 & $(1, 0)\times (1, 0)\times (2, 0)$& \multicolumn{8}{c|}{$x_A = \frac{14}{5}x_B = 2x_C = 7x_D$}\\
			4 & 2 & $(0, -1)\times (0, 1)\times (2, 0)$& \multicolumn{8}{c|}{$\beta^g_1=-3$, $\beta^g_4=1$}\\
			& & & \multicolumn{8}{c|}{$\chi_1=\sqrt{5}$, $\chi_2=\frac{7}{\sqrt{5}}$, $\chi_3=\frac{4}{\sqrt{5}}$}\\
			\hline
		\end{tabular}
	\end{center}
\end{table}

\section{String-scale Gauge Coupling Unification}
Once we obtain the full list of supersymmetric Pati-Salam models, we can discuss  the gauge coupling relations at the string scale.
As shown in Table~\ref{tb:model1}, Model 1 has exact gauge coupling unification at GUT scale from brane construction, while the other models do not.

Using the current precision electroweak data and setting $M_{\text{SUSY}}=3$ TeV, we study the gauge coupling relations
at string scale~(around $5\times10^{17}$ GeV ) 
by solving two-loop renormalization group equations (RGEs)~\cite{Barger:2005qy,Barger:2007qb}, i.e., $M_U\simeq M_{\text{string}}$. 
The generic gauge coupling relations at string scale for intersecting D6-brane models 
can be written as $g_a^2=k_2\,g_b^2=k_Y\,g_Y^2=g_{U}^2$, where 
$k_Y$ and $k_2$ are constants for each model. For simplicity, we call these gauge coupling relations as
gauge coupling unification by redefining  
$\alpha_1 \equiv k_Yg_Y^2/4\pi$, $\alpha_2 \equiv k_2g_b^2/4\pi$, and $\alpha_3 \equiv g_a^2/4\pi$.
To define the unification scale, 
we choose the evolution under the conditions $\alpha_{U}^{-1}\equiv \alpha_1^{-1}=(\alpha_2^{-1}+\alpha_3^{-1})/2$ and  $\Delta=|\alpha_1^{-1}-\alpha_2^{-1}|/\alpha_1^{-1}$ as in Refs.~\cite{Chen:2017rpn,Chen:2018ucf}.
In our study, the difference between $\alpha_{\text{string}}^{-1}$ and $\alpha_2^{-1}$ or $\alpha_3^{-1}$ is limited to be less than $1.0\%$.

\begin{figure}[h]
	\centering
	\includegraphics{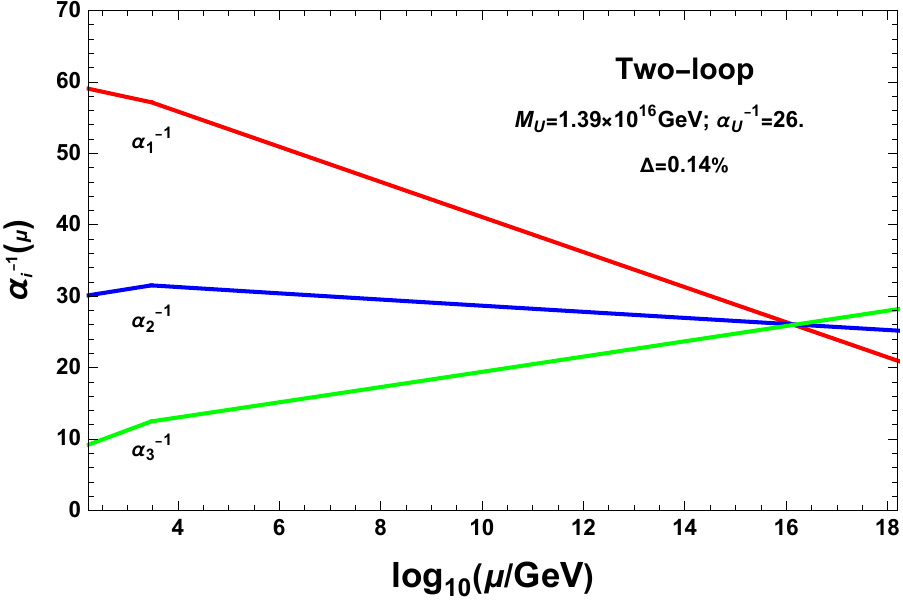}
	\caption{Gauge coupling unification at  $1.39\times 10^{16}$ GeV in Model 1}
	\label{fig:M1}
\end{figure}

For Model 1, the canonical hypercharge normalization $k_Y=5/3$ and $k_2=1$. Utilizing two-loop RGEs method, we obtain that the gauge coupling unification can be achieved at  $1.39\times 10^{16}$ GeV which is just the supersymmetric GUT scale value as shown in~Fig.\ref{fig:M1}. 
Note that this unification scale is about one order magnitude smaller than string scale $5\times10^{17}$ GeV.

In~\cite{Blumenhagen:2003jy}, it was presented that additional 
vector-like matter can push the unification at GUT scale up to
Planck scale for $SU(5)$ and $SO(10)$ GUT theories. 
Instead of studying pushing up the unification scale with additional vector-like matter,  we will study the feature while introducing vector like particle to the models without GUT scale unification, such as Model 2 and 3. 
To achieve canonical gauge coupling unification, we first introduce
the vector-like particles whose quantum numbers are the same as those
of the SM fermions and their Hermitian conjugate, see~\cite{Barger:2007qb} and references therein. 
Their quantum numbers under 
$SU(3)_C \times SU(2)_L \times U(1)_Y$ and their
contributions to one-loop beta functions, $\Delta b \equiv (\Delta
b_1, \Delta b_2, \Delta b_3)$ as complete supermultiplets are
\begin{eqnarray}
	&& XQ + {\overline{XQ}} = {\mathbf{(3, 2, {1\over 6}) + ({\bar 3}, 2,
			-{1\over 6})}}\,, \quad \Delta b =({1\over 5}, 3, 2)\,;\\ 
	&& XU + {\overline{XU}} = {\mathbf{ ({3},
			1, {2\over 3}) + ({\bar 3},  1, -{2\over 3})}}\,, \quad \Delta b =
	({8\over 5}, 0, 1)\,;\\ 
	&& XD + {\overline{XD}} = {\mathbf{ ({3},
			1, -{1\over 3}) + ({\bar 3},  1, {1\over 3})}}\,, \quad \Delta b =
	({2\over 5}, 0, 1)\,;\\  
	&& XL + {\overline{XL}} = {\mathbf{(1,  2, {1\over 2}) + ({1},  2,
			-{1\over 2})}}\,, \quad \Delta b = ({3\over 5}, 1, 0)\,;\\ 
	&& XE + {\overline{XE}} = {\mathbf{({1},  1, {1}) + ({1},  1,
			-{1})}}\,, \quad \Delta b = ({6\over 5}, 0, 0)\,;\\ 
	&& XG = {\mathbf{({8}, 1, 0)}}\,, \quad \Delta b = (0, 0, 3)\,;\\ 
	&& XW = {\mathbf{({1}, 3, 0)}}\,, \quad \Delta b = (0, 2, 0)\,;\\
	&& XT + {\overline{XT}} = {\mathbf{(1, 3, 1) + (1, 3,
			-1)}}\,, \quad \Delta b =({{18}\over 5}, 4, 0)\,;\\ 
	&& XS + {\overline{XS}} = {\mathbf{(6,  1, -{2\over 3}) + ({\bar 6},
			1, {2\over 3})}}\,, \quad \Delta b = ({16\over 5}, 0, 5)\,;\\ 
	&& XY + {\overline{XY}} = {\mathbf{(3, 2, -{5\over 6}) + ({\bar 3}, 2,
			{5\over 6})}}\,, \quad \Delta b =(5, 3, 2)\,.\,
\end{eqnarray}

\begin{figure}[t]
	\centering
	{\includegraphics{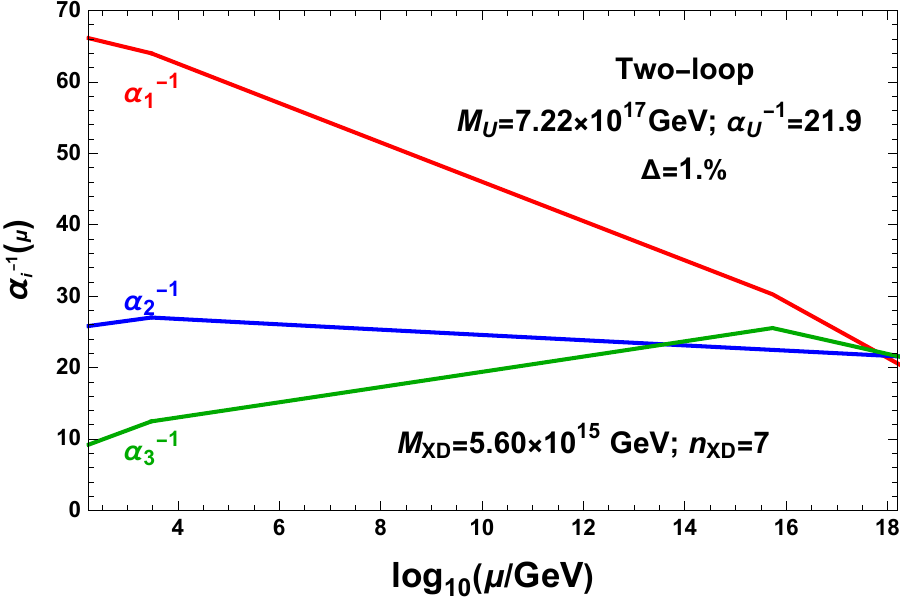}}
	\caption{Gauge coupling unification at  $7.22\times 10^{17}$ GeV in Model 2}\label{fig:RGE-M2}
\end{figure}

For Models 2 as shown in Table~\ref{tb:model2}, the gauge coupling relation at GUT scale is $g^2_a=\frac{7}{6}\,g^2_b=\frac{5}{6}\, g^2_c=\frac{25}{28}\,(\frac{5}{3}g^2_Y)=\frac{8}{27}\, 5^{3/4} \sqrt{7} \pi  e^{\phi ^4}$, without gauge coupling unification from brane construction.  
By introducing additional pairs vector-like particles ($XD$, $\overline{XD}$) for Model 2 with $k_Y=125/84,~k_2=7/6$ with $\Delta b = ({2\over 5}, 0, 1)$, it would be possible to ``bend'' the values of $\alpha_1^{-1}, \alpha_3^{-1}$ to achieve gauge coupling as shown in Figure~\ref{fig:RGE-M2}. 
Here the number of additional particles is important as it determines the slope of bending and, as follows, the energy scale of the gauge unification. 
We speculate the number given by brane intersection might lead to string scale gauge coupling unification. 
Recall the quantum numbers of ($XD$, $\overline{XD}$) under $SU(3)_C\times SU(2)_L \times U(1)_Y$ are $XD=(3,1,-1/3)$, $\overline{XD}=(\bar{3},1,1/3)$. It is obvious that these vector-like particles arise from $a$ and $c$ stack of brane intersection, representing $SU(3)_C$ and $U(1)_Y$ respectively.  

Read off the wrapping numbers $(n_x^i, l_x^i)$ for $a$ and $c$ stack of the from Table~\ref{tb:model2}, we note that $(n_a^3, l_a^3)$ and  $(n_c^3, l_c^3)$ has the same value yet with opposite sign. This indicates the D6-branes wrapping on the third torus are parallel to each other. Therefore there is no intersection on the third torus, but only on the first and second torus.  Following Eq.~\eqref{intersection}, we have intersection number
$I_{ac}=2^{-k}\prod_{i=1}^2(n_a^i\,l_c^i-n_c^i\,l_a^i)= 7$.  This indicates there are $7$ number of such vector-like particles naturally arise from brane construction.
By introducing $7$ pairs of vector-like particle ($XD$, $\overline{XD}$) for Model 2 with $k_Y=125/84,~k_2=7/6$, via two-loop RGEs method, we indeed obtain string scale gauge coupling unification at around  $7.22\times 10^{17}$ GeV  presented in Figure~\ref{fig:RGE-M2}. 

For Models 3 as shown in Table~\ref{tb:model3}, the gauge coupling relation at GUT scale is $g^2_a=\frac{5}{6}\,g^2_b=\frac{7}{6}\,g^2_c=\frac{35}{32}\,(\frac{5}{3}g^2_Y)=\frac{8}{27}\, 5^{3/4} \,\sqrt{7} \pi  e^{\phi ^4}$, again without gauge coupling unification from brane construction. Similar as for Model 2, we find vector-like particle  ($XQ$, $\overline{XQ}$) with quantum number 
$XQ=(3,2,1/6)$ and $\overline{XQ}=(\bar{3},2,-1/6)$ are nature candidate to realize string scale gauge coupling unification.  These particles are realized from $a$ and $b$ stacks of brane intersection, with intersection number $I_{ab}=2^{-k}\prod_{i=1}^2(n_a^i\,l_b^i-n_b^i\,l_a^i)= 7$. 

\begin{figure}[t]
	\centering
	{\includegraphics{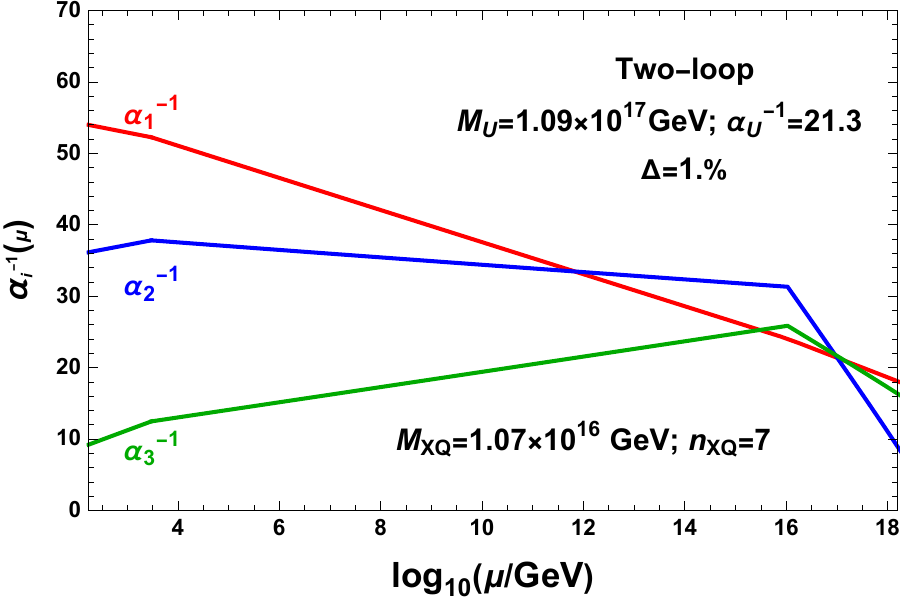}}
	\caption{Gauge coupling unification at $1.09\times 10^{17}$ GeV in Model 3}\label{fig:RGE-M3}
\end{figure}

Thus, by introducing $7$ pairs of vector-like particle ($XQ$, $\overline{XQ}$) in Model 3 with $k_Y=175/96,~k_2=5/6$, we  achieve gauge coupling unification at around string scale $1.09\times 10^{17}$ GeV, which we show in Figure~\ref{fig:RGE-M3}. 
The masses of the extra vector-like particles for Model 2 and Model 3 are also presented in Figure~\ref{fig:RGE-M2} and Figure~\ref{fig:RGE-M3}, respectively. The right-side kinked points on  the plots indicate masses of the extra vector-like particles on the $x$-axis.

One may ask the reason why only these type of vector-like particle are turned on, but not the other type of  vector-like particles. This is realized by setting the minimal distance squared $Z^2_{(ac')}$
(in $1/M_s$ units) between parallel D6-branes along the third torus is small, {\it i.e.},  the minimal length squared of the stretched string is small.
Therefore, we have the light scalars with
squared-masses $Z^2_{(ab')}/(4\pi^2 \alpha')$ from the NS sector,
and the light fermions with the same masses from the R
sector~\cite{Uranga, imr}. 
These scalars and fermions form the
4-dimensional $N=2$ hypermultiplets, so, \emph{e.g.} for Model 2 we obtain the
$I_{ac}$ (the intersection numbers for $a$ and $c$ stacks
on the first two two-tori) vector-pairs of
($XD$, $\overline{XD}$) under $SU(3)_C\times SU(2)_L \times U(1)_Y$ with quantum number $XD=(3,1,-1/3)$, $\overline{XD}=(\bar{3},1,1/3)$.
This setting applies to Model 3 with ${\cal N}=2$ vector-like particles ($XQ$, $\overline{XQ}$) also.

\section{Conclusions}

In~\cite{Li:2021-2},  we obtain all the three-family ${\cal N}=1$ supersymmetric 
Pati-Salam models from intersecting D6-branes through systematic method. 
The gauge symmetry can be broken down to the SM via the D6-brane splitting 
and supersymmetry preserving Higgs mechanism. There are only 33 independent models with different gauge coupling relations at string scale in total after modding out the equivalent relations.
In particular, one and only one model has natural gauge coupling unification at GUT scale through brane construction. 

In this paper, we study the gauge coupling unification at string scale for the models without direct gauge unification at GUT scale from brane construction. 
Taking models with approximate gauge coupling unification at GUT scale as example,  we showed that 
 gauge coupling unification can be realized at string scale by introducing 
pairs of vector-like particles from ${\cal N}=2$ sector.
In particular, to achieve the gauge unification as string scale, we propose that the number of these introduced vector-like particles are fully determined by the brane intersection numbers while there are two D6-brane parallel to each other along one two-torus.
Intriguingly, with two-loop RGEs method,  we explicitly showed for Model 2 and 3 that  gauge coupling unified at string-scale $10^{17}$ GeV by introducing $7$ pairs vector-like particles ($XD$, $\overline{XD}$) and ($XQ$, $\overline{XQ}$) with their masses given respectively. 

We expect that by introducing pairs of vector-like particles from  ${\cal N}=2$ sector for intersecting brane models, gauge coupling unification at string scale can be achieved for generic D-brane models. In particular, the number of vector-like particles are fully determined by brane intersection numbers. 


\begin{acknowledgments}

TL is supported by the National Key Research and Development Program of China Grant No. 2020YFC2201504, as well as by the Projects No. 11875062 and No. 11947302 supported by the National Natural Science Foundation of China.
 WH is supported by KIAS Individual Grant MG080401. RS is supported by KIAS Individual Grant PG080701.  LW is supported by Projects 2020JQ-804 from the Natural Science Basic Research Plan  in Shaanxi Province of China and 20JK0685 from Shaanxi Provincial Education Department.  

\end{acknowledgments}



\begin{thebibliography}{99} 
	
	\bibitem{lpt}
	J.~Lykken, E.~Poppitz and S.~P.~Trivedi, Nucl.\ Phys.\ B {\bf
		543}, 105 (1999).
	
	
	\bibitem{magnetised}
	C.~Angelantonj, I.~Antoniadis, E.~Dudas and A.~Sagnotti, Phys.
	Lett. B {\bf 489} (2000) 223.
	
	
	\bibitem{MCJW}
	M.~Cveti\v c, M.~Pl\"umacher and J.~Wang, JHEP {\bf 0004}, 004 (2000);
	M.~Cveti\v c, A.~M.~Uranga and J.~Wang, Nucl.\ Phys.\ B {\bf 595}, 63
	(2001).
	
	\bibitem{Ibanez}
	G.~Aldazabal, A.~Font, L.~E.~Ib\'a\~nez and G.~Violero, Nucl.\ Phys.\
	B {\bf 536}, 29 (1998); G.~Aldazabal, L.~E.~Ib\'a\~nez, F.~Quevedo and
	A.~M.~Uranga, JHEP {\bf 0008}, 002 (2000).
	
	\bibitem{MKRR}
	M.~Klein and R.~Rabadan, JHEP {\bf 0010}, 049 (2000).
	
	
	\bibitem{urangac}
	J.~F.~G.~Cascales and A.~M.~Uranga,
	hep-th/0311250.
	
	\bibitem{afiru}
	G.~Aldazabal, S.~Franco, L.~E.~Ib\'a\~nez, R.~Rabad\'an and
	A.~M.~Uranga, JHEP {\bf 0102}, 047 (2001).
	
	\bibitem{Uranga}
	G.~Aldazabal, S.~Franco, L.~E.~Ib\'a\~nez, R.~Rabadan and
	A.~M.~Uranga, J.\ Math.\ Phys.\  {\bf 42}, 3103 (2001).
	
	\bibitem{imr}
	L.~E.~Ib\'a\~nez, F.~Marchesano and R.~Rabad\'an, JHEP {\bf 0111},
	002 (2001).
	
		\bibitem{bgkl}
	R.~Blumenhagen, L.~G\"orlich, B.~K\"ors and D.~L\"ust, JHEP {\bf
		0010}, 006 (2000).
		
	\bibitem{bkl}
	R.~Blumenhagen, B.~K\"ors and D.~L\"ust, JHEP {\bf 0102}, 030 (2001).
	
	
	\bibitem{CSU2}
	M.~Cveti\v c, G.~Shiu and A.~M.~Uranga, Nucl.\ Phys.\ B {\bf 615},
	3 (2001).
	
	\bibitem{Blumenhagen:2003jy} 
	R.~Blumenhagen, D.~L\"ust, S.~Stieberger,
	JHEP {\bf 07}, 036 (2003).
	

	\bibitem{CPS}
	M.~Cveti\v c, I.~Papadimitriou and G.~Shiu,
	Nucl.\ Phys.\ B {\bf 659}, 193 (2003)
	Erratum: [Nucl.\ Phys.\ B {\bf 696}, 298 (2004)].
	
	\bibitem{Cvetic:2004ui} 
	M.~Cveti\v c, T.~Li and T.~Liu,
	Nucl.\ Phys.\ B {\bf 698}, 163 (2004).
	
	\bibitem{Blumenhagen:2005mu} 
	R.~Blumenhagen, M.~Cvetic, P.~Langacker and G.~Shiu,
	Ann.\ Rev.\ Nucl.\ Part.\ Sci.\  {\bf 55}, 71 (2005).
	
	\bibitem{Blumenhagen:2006ci} 
	R.~Blumenhagen, B.~Kors, D.~L\"ust, S.~Stieberger,
	Phys.\ Rept. \ {\bf 445}, 1-193 (2007).
	
	



\bibitem{CLW}
M.~Cveti\v c, P.~Langacker and J.~Wang, Phys.\ Rev.\ D {\bf 68},
046002 (2003).

\bibitem{blumrecent}
R.~Blumenhagen, L.~Gorlich and T.~Ott,
JHEP {\bf 0301}, 021 (2003).

\bibitem{Honecker}
G.~Honecker,
Nucl.\ Phys.\ B {\bf 666}, 175 (2003).

\bibitem{Cvetic:2004nk} 
M.~Cveti\v c, P.~Langacker, T.~Li and T.~Liu,
Nucl.\ Phys.\ B {\bf 709}, 241 (2005).



\bibitem{Chen:2005mj} 
C.~M.~Chen, T.~Li and D.~V.~Nanopoulos,
Nucl.\ Phys.\ B {\bf 732}, 224 (2006).




\bibitem{Chen:2007px} 
C.~M.~Chen, T.~Li, V.~E.~Mayes and D.~V.~Nanopoulos,
Phys.\ Lett.\ B {\bf 665}, 267 (2008).


\bibitem{Chen:2007zu} 
C.~M.~Chen, T.~Li, V.~E.~Mayes and D.~V.~Nanopoulos,
Phys.\ Rev.\ D {\bf 77}, 125023 (2008).


\bibitem{Li:2019nvi}
T.~Li, A.~Mansha and R.~Sun, EPJC {\bf 81}, 82 (2021).

\bibitem{Li:2021}
T.~Li, A.~Mansha, R.~Sun, L.~Wu, and W.~He, Phys.Rev.D {\bf 104}, 046018 (2021).

\bibitem{Jim:brane}
J.~Halverson, B.~Nelson, and F.~Ruehle, JHEP {\bf  06}, 003 (2019).

\bibitem{Loges:2021hvn}
G. J.~Loges, G.~Shiu, Fortschritte der Physik 2022, 2200038,
arXiv: 2112.08391[hep-th].


\bibitem{LMSW}
M.~Sabir, T.~Li, A.~Mansha, and X. C. Wang, in preparation.

\bibitem{Li:2021-2}
T.~Li, R.~Sun, and W.~He, arXiv: 2112.09632 [hep-th].

\bibitem{Li:2021-3}
T.~Li, R.~Sun, and L.~Wu, in preparation.

\bibitem{bdl}
M.~Berkooz, M.~R.~Douglas and R.~G.~Leigh, Nucl. Phys. B {\bf 480}, 265
(1996).




\bibitem{Blumenhagen:2006xt}
R.~Blumenhagen, M.~Cvetic and T.~Weigand,
Nucl. Phys. B \textbf{771}, 113 (2007).

\bibitem{Haack:2006cy}
M.~Haack, D.~Krefl, D.~Lust, A.~Van Proeyen and M.~Zagermann,
JHEP \textbf{01}, 078 (2007).


\bibitem{Florea:2006si}
B.~Florea, S.~Kachru, J.~McGreevy and N.~Saulina,
JHEP \textbf{05}, 024 (2007).







\bibitem{Barger:2005qy} 
V.~Barger, J.~Jiang, P.~Langacker and T.~Li,
Nucl.\ Phys.\ B {\bf 726}, 149 (2005).

\bibitem{Barger:2007qb}
V.~Barger, N.~G.~Deshpande, J.~Jiang, P.~Langacker and T.~Li,
Nucl. Phys. B \textbf{793}, 307 (2008).


\bibitem{Chen:2017rpn}
H.~Y.~Chen, I.~Gogoladze, S.~Hu, T.~Li and L.~Wu,
Eur. Phys. J. C \textbf{78}, 26 (2018).

\bibitem{Chen:2018ucf}
H.~Y.~Chen, I.~Gogoladze, S.~Hu, T.~Li and L.~Wu,
Int. J. Mod. Phys. A \textbf{35}, 2050117 (2020). 





\end{thebibliography}

\end{document}